\documentclass{article}

\usepackage{spconf}
\usepackage{graphicx}
\usepackage{float}
\usepackage{epsfig}
\usepackage{amsmath}
\usepackage{amssymb}
\usepackage{bm}
\usepackage{color}
\usepackage{url}
\usepackage{hyperref}
\hypersetup{
    colorlinks=true,
    citecolor=blue,
    urlcolor=blue,
    linkcolor=blue
}

\title{Rate Distortion Optimization over Large Scale Video Corpus with Machine Learning}
\name{Sam John, Akshay Gadde and Balu Adsumilli}
\address{
Google, Mountain View, CA \\
\{samjohn, agadde, badsumilli\}@google.com
}
\begin{document}
%
\maketitle
\begin{abstract}
We present an efficient codec-agnostic method for bitrate allocation over a large scale video corpus with the goal of minimizing the average bitrate subject to constraints on average and minimum quality. Our method clusters the videos in the corpus such that videos within one cluster have similar rate-distortion (R-D) characteristics. We train a support vector machine classifier to predict the R-D cluster of a video using simple video complexity features that are computationally easy to obtain. The model allows us to classify a large sample of the corpus in order to estimate the distribution of the number of videos in each of the clusters. We use this distribution to find the optimal encoder operating point for each R-D cluster. Experiments with AV1 encoder show that our method can achieve the same average quality over the corpus with $22\%$ less average bitrate.   
\end{abstract}
\begin{keywords}
rate distortion optimization, clustering, machine learning, adaptive streaming, YouTube
\end{keywords}
\vspace{-0.5\baselineskip}
\section{Introduction}
\label{sec:intro}
\vspace{-0.5\baselineskip}
For massive video streaming platforms such as YouTube, it is desirable to deliver the best video quality with minimum bitrates since it directly contributes to the streaming cost for these platforms as well to the data cost and quality of experience of the users. These platforms have various types of video content ranging from song lyrics and simple animations to sports and gaming~\cite{wang-mmsp-19}. For a given video encoder, we can compute the rate-distortion (R-D) curve of a video by encoding the video at different bitrates and plotting the distortion achieved for each bitrate. The R-D characteristics for each content type is significantly different than others. Even within one video, R-D characteristics can vary significantly over time. A video can be divided into chunks and a separate R-D curve can be obtained for each chunk. Given the R-D curves for each chunk in a video corpus, we can define a problem of finding the optimal bitrate for encoding each chunk such that an aggregate measure of distortion is minimized subject to a constraint on average bitrate (or average bitrate is minimized subject to a constraint on quality). Solving such an optimization problem directly at a large scale such as YouTube's, which contains millions of hours of videos~\cite{yt-for-press}, is infeasible since it would involve encoding each chunk at multiple bitrates to get the R-D curve and then solving a non-linear optimization problem in millions of variable. Moreover, it is necessary to encode each video chunk into multiple representations at different bitrates and resolutions for adaptive streaming over networks of varying bandwidth~\cite{seufert-cst-15}. This introduces additional complexity to the problem since we need to find the optimal bitrates for all representations. 

The problem of R-D optimal bitrate allocation over coding units given their R-D characteristics has been studied before~\cite{ortega-spm-98}. The authors show that the optimal bitrate allocation is such that the marginal gain in quality achieved by spending one extra unit of bitrate is equal for all coding units. However, they do not consider bitrate allocation over different representations of a coding unit. Toni et al.~\cite{toni-tomm-15} consider the problem of selecting optimal representation of a video for adaptive streaming taking into account network dynamics. The problem of finding optimal bitrates for different representations of a video chunk subject to constraint on average delivered quality considering the distribution of user bandwidths and viewports is studied by Chen et al.~\cite{chen-icip-19}. All of these methods assume that the R-D curves of all coding video chunks in the corpus are known and that the number of encoding bitrates to optimize is small. These assumptions are not feasible for R-D optimization over a large scale video corpus.

\begin{figure}
\centering
\includegraphics[scale=0.60]{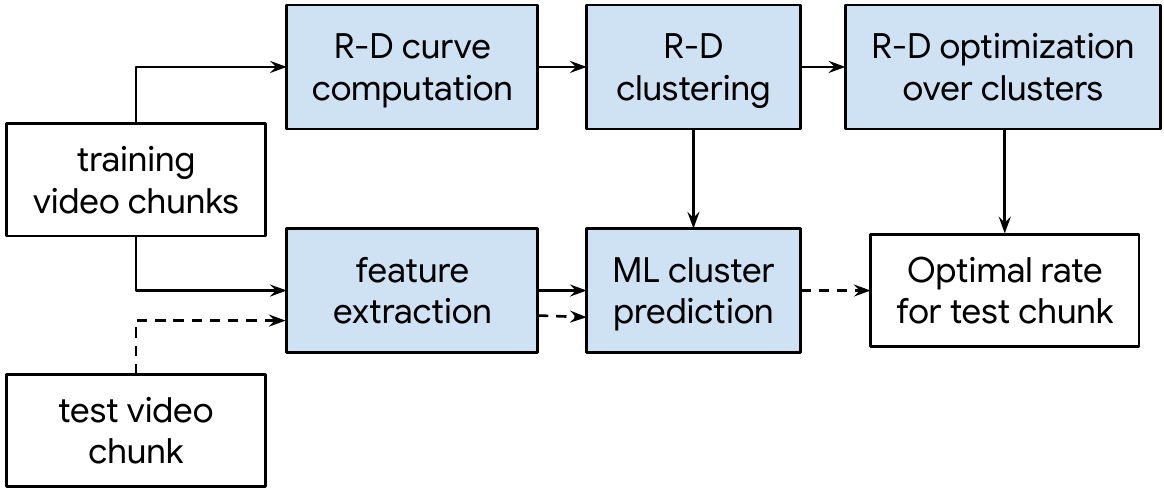}
\vspace{-0.5\baselineskip}
\caption{Outline of the proposed algorithm. White boxes show the input and output data. Solid arrows denote the training flow and dotted arrows indicate how models are used.}
\label{fig:algo-summary}
\vspace{-0.5\baselineskip}
\end{figure}

We propose a new efficient codec-agnostic method for allocating bitrates for all video chunks in a large scale corpus with the goal of minimizing the average bitrate while maintaining aggregate quality. Our method does not require encoding of all chunks in the corpus at multiple bitrates to get their R-D curves. This reduces the computational complexity significantly, especially at scale. Instead we use simple video complexity features obtained from encoder pass-log to predict the R-D curve of a chunk using machine learning. We cluster the video chunks in the corpus into multiple categories based on their approximate R-D curves. We demonstrate that the number of clusters required to model the variation in the R-D characteristics across the corpus is much less than the size of the corpus. Because all video chunks in a category have similar R-D characteristics, we can optimize for bitrates over categories instead of individual chunks. This requires solving a relatively small non-linear optimization problem with dimensionality equal to the number of clusters. 
The proposed rate allocation method is outlined in Figure~\ref{fig:algo-summary}.

The rest of the paper is organized as follows. In Section~\ref{sec:ml}, we describe the proposed R-D curve prediction method based on R-D curve clustering and video classification using video complexity features. Section~\ref{sec:opt} explains our R-D optimization formulation to compute the optimal bitrates for all video categories. Experimental results in Section~\ref{sec:experiment} show the efficacy our proposed approach. We conclude the paper in Section~\ref{sec:conclusion}.

\section{R-D Curve Modeling}
\label{sec:ml}
\vspace{-0.5\baselineskip}
Our method attempts to categorize the video chunks in the corpus such that all video chunks in one category have similar R-D curves. We do this using a two step approach (see Figure~\ref{fig:algo-summary}). In the first step, we cluster the R-D curves of training video chunks that are randomly sampled from the corpus. In the second step, we build a classification model to predict which R-D cluster a chunk belongs to using simple video complexity features obtained by fast one-pass analysis of the chunk with an encoder (i.e., encoder pass-log). R-D optimization is done over these clusters using their centroid R-D curves.
\vspace{-0.5\baselineskip}
\subsection{Clustering Video Corpus Based on R-D Curves}
\label{sec:clustering}
Let $n$ be the number of training video chunks selected from the corpus. We obtain $s$ points on the R-D curve of each training video chunk by encoding the chunk at fixed encoder operating points, $[q_1, \ldots, q_{s}]$. Each operating point $q_j$ corresponds to a quantization parameter (QP) or a constant rate factor (CRF) used by the encoder. Encoding a chunk $i$ at operating point $q_j$ results in a representation with bitrate $r_j$ and distortion $d_j$ for a chunk. The R-D curve samples for chunk $i$ can then be represented by a vector $\mathbf{x}_i \in \mathbb{R}^{2s}$ of rate-distortion values, $[r_1, \ldots, r_s, d_1, \ldots, d_s]$.
We normalize each component of $\mathbf{x}$ as $x_j^{\text{norm}} = (x_j - m_j)/\sigma_j$, where $m_j$ and $\sigma_j$ are sample mean and sample standard deviation of $x_j$ respectively. We cluster the vectors $\mathbf{x}_1^{\text{norm}}, \ldots, \mathbf{x}_{n}^{\text{norm}}$ into $k$ ($\ll n$) clusters $\mathbf{C} = \{C_1, \ldots, C_k\}$ using $k$-means.
$L^2$ distance between normalized R-D points is used to define the cost function for clustering. It is reasonable to normalize and compute the distances for each component of $\mathbf{x}$ across different video chunks since each component corresponds to a fixed encoder operating point.
We use the centroid $\bm{\mu}_l$ of cluster $C_l$ to get curves $\rho_l(q)$ and $\delta_l(q)$ for mapping an operating point $q$ to bitrate and distortion values respectively. These centroid curves are expected to be a good approximation for bitrate $r_i(q)$ and distortion $d_i(q)$ for any video chunk $i \in C_l$.
%

\begin{figure}
\centering
\includegraphics[scale=0.40]{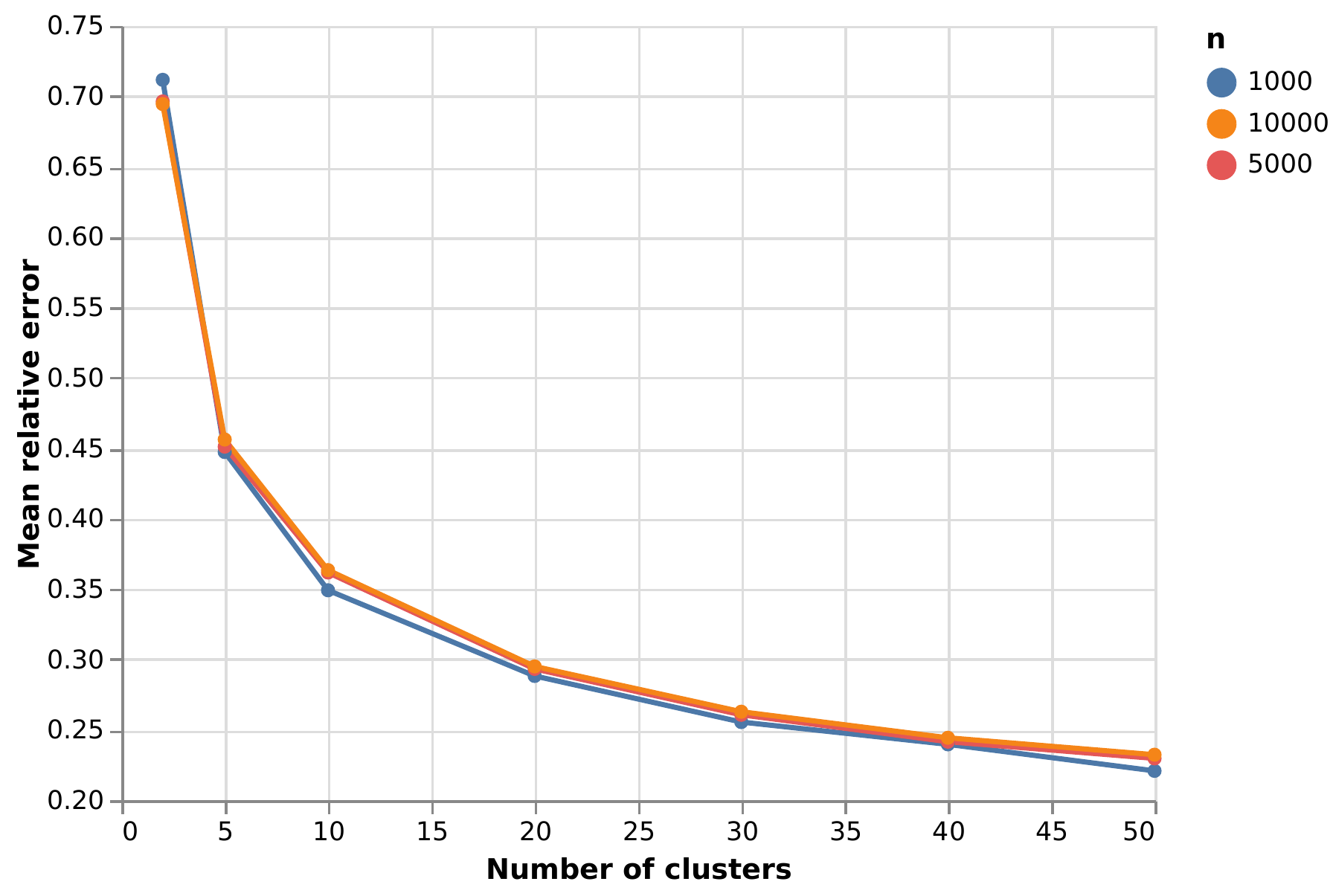}
\vspace{-0.5\baselineskip}
\caption{Plot of mean relative error between the training R-D points and corresponding cluster centroids vs. number of clusters for different number of training R-D points.}
\label{fig:num-clusters}
\vspace{-1.0\baselineskip}
\end{figure}
\begin{figure*}
\centering
\includegraphics[scale=0.36]{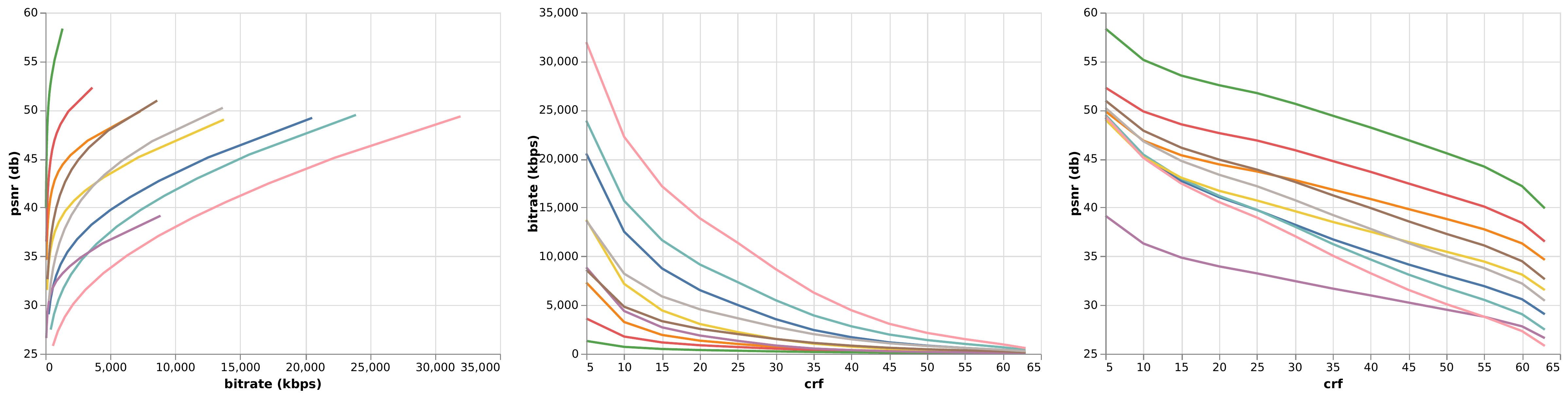}
\vspace{-0.5\baselineskip}
\caption{Plots of PSNR vs. bitrate, bitrate vs. CRF and PSNR vs. CRF for different cluster centroids}
\label{fig:rd-centroids}
\vspace{-0.5\baselineskip}
\end{figure*}

%
The number of clusters, $k$, needed to capture all the variation in the R-D characteristics in the corpus is determined empirically. Figure~\ref{fig:num-clusters} shows the mean relative error between the R-D points for a corpus sample of $n$ chunks and their corresponding cluster centroids (i.e., $\frac{\sum_i \|\mathbf{x}_i^{\text{norm}} - \bm{\mu}_l\|}{\sum_i \|\mathbf{x}_i^{\text{norm}}\|}$) as function of $k$ for different values on $n$. The data points are obtained by encoding video chunks at different CRF values with AV1 encoder~\cite{chen-pcs-18}. It shows that approximating the corpus R-D points by the corresponding cluster centroids results in a small relative error that reduces slowly as the number of clusters exceeds $10$. Moreover, the number of clusters needed to achieve this small error does not increase with the sample size $n$. 
Figure~\ref{fig:rd-centroids} show the plots of bitrate vs. CRF, distortion vs. CRF and distortion vs. bitrate for different cluster centroids for $k = 10$. These plots show that the marginal reduction in distortion achieved by allocating higher bitrate varies significantly across clusters. Therefore, the optimal operating point for achieving the best rate-distortion tradeoff will be different for each cluster.
\vspace{-0.5\baselineskip}
\subsection{Predicting the R-D Cluster for a Video}
\label{sec:classification}
It is not feasible to sample the R-D curve for every video chunk in the corpus in order to determine its R-D cluster $C_l$ since it involves encoding each chunk multiple times at different operating points.
In order to circumvent this problem, we train a support vector machine (SVM) classifier to predict the R-D cluster of a chunk using video complexity features that are computationally much cheaper to obtain. 

\begin{figure}
    \centering
    \includegraphics[scale=0.36]{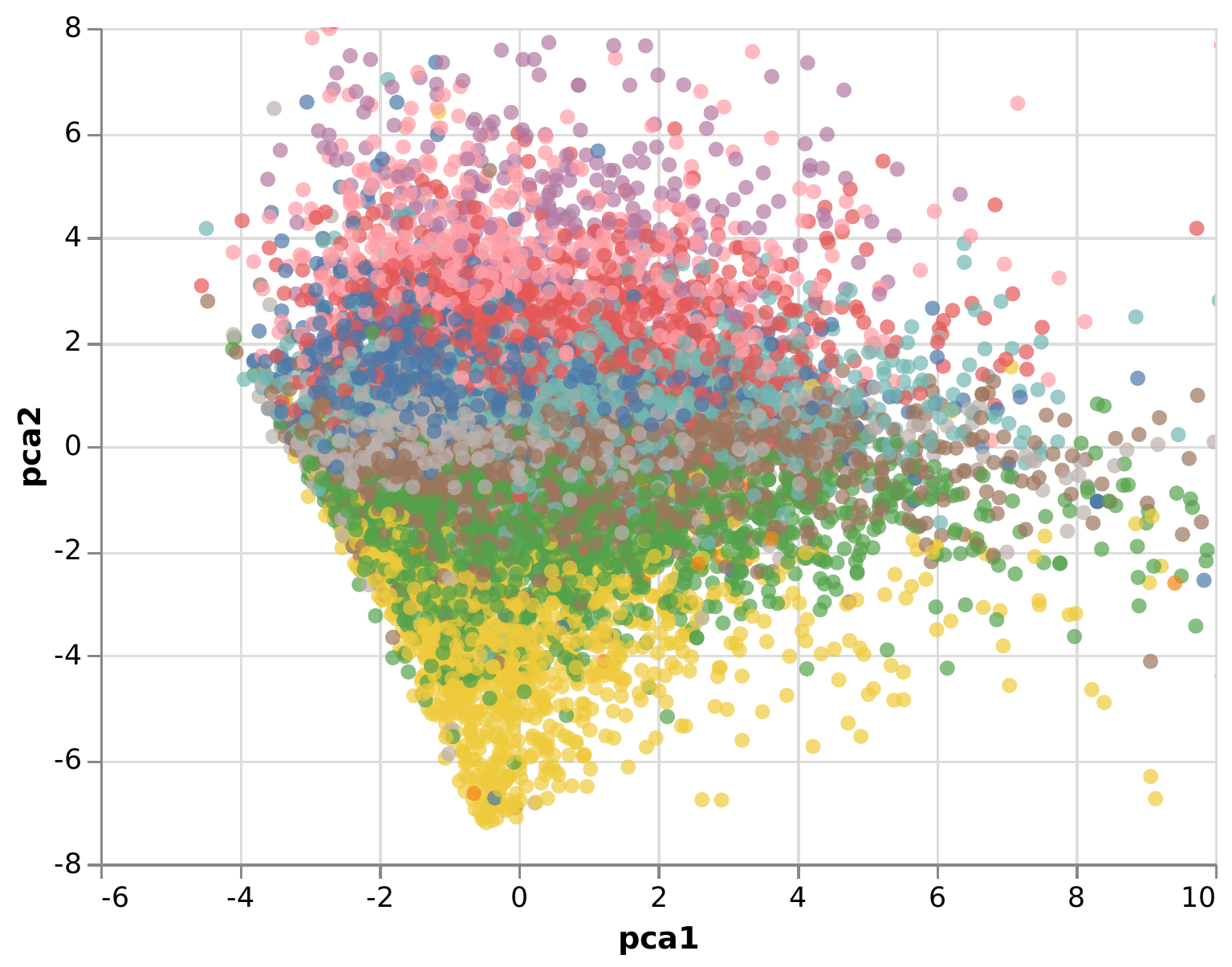}
    \vspace{-0.5\baselineskip}
    \caption{Projection of the training features using first two principal components. The color denotes the R-D class of a point. R-D classes exhibit some clustering in this space.}
    \label{fig:pca_clusters}
    \vspace{-1.0\baselineskip}
\end{figure}

Specifically, we use $22$ of the passlog features given by the AV1 encoder. AV1 encoder generates this passlog by doing an analysis pass over a video without fully encoding it. The features include statistics related to prediction modes (inter or intra), prediction errors, reference frames used for inter prediction and motion vectors for each frame in the video~\cite{libaom-repo}. These features are a good indicator of the spatial and temporal complexity of a video. Therefore, they are useful for predicting the R-D characteristics of the video (see Figure~\ref{fig:pca_clusters}).
In order to train the classifier, the ground truth class labels for video chunks in the training set are obtained by clustering their R-D curves.
Therefore, centroid R-D curve of the predicted cluster of a video is expected to be good approximation for the R-D curve of the video.
%

%
%
The idea of clustering the R-D curves and building a model for R-D cluster prediction was proposed independently by Ling et al.~\cite{ling-ei-20}. However, our method and the method in \cite{ling-ei-20} have some key differences. Firstly, the method in \cite{ling-ei-20} uses the Bjontegaard Delta (BD) rate between the PSNR vs. bitrate curves as the distance metric in clustering. The problem with using this metric is that two R-D curves with substantially different slopes may have a very small BD rate distance. However, taking these differences in slopes into account is critical for efficient rate utilization. Ling et al. also use different features for predicting the R-D curve cluster. We find that using the encoder pass log features is computationally efficient and allows accurate R-D cluster prediction.  

\section{R-D Optimization over Video Corpus}
\label{sec:opt}
\vspace{-0.5\baselineskip}
Our goal is to find the optimal encoder operating points for all R-D clusters so that the average corpus bitrate is minimized while the average and worst-case distortions remain below certain thresholds. 
In order to compute these averages, we need to estimate the distribution of the number of video chunks in different clusters. 
%
We do this by classifying a large sample of the video corpus into different clusters using the SVM model proposed in Section~\ref{sec:classification}. 
%
We can use this distribution to compute the average bitrate and distortion for the corpus for a given set of operating points for the clusters based on their centroid R-D curves. 

Let $\mathbf{q} = [q_1, \ldots, q_k]$ be the encoder operating points for clusters $C_1, \ldots, C_k$ respectively. Let $\rho_l(q_l)$ and $\delta_l(q_l)$ denote the bitrate and distortion obtained by encoding a video chunk in cluster $C_l$ at operating point $q_l$ as given by the centroid R-D curves for $C_l$. The optimal value of $\mathbf{q}$ is defined as the solution to the following problem:
\begin{align}
\underset{\mathbf{q}}{\text{minimize }}& \sum_{l = 1}^{k} w_l \rho_l(q_l)  \nonumber \\
\text{subject to }& \sum_{l = 1}^{k} w_l\delta_l(q_l) \leq D_{\text{avg}} \nonumber \\
& \max_{l \in \{1, \ldots, k\}} \delta_l(q_l) \leq D_\text{max},
\label{eq:opt}
\end{align}
where $w_l$ is the fraction of video chunks in cluster $C_l$. Note that $\sum_l w_l = 1$. The solution to the above problem, $\mathbf{q}^\star$, will minimize the total bitrate while maintaining the given constraints on distortion. Any video chunk in cluster $C_l$ is encoded using the optimal operating point $q_l^\star$ for that cluster. 
%


\section{Experiments}
\label{sec:experiment}
\vspace{-0.5\baselineskip}
In order to evaluate the performance of the proposed method, we use a set of $n=14000$ videos at $480$p resolution randomly sampled from the YouTube corpus. We sample a 5 second long chunk from each video. Using short chunks ensures that the R-D characteristics do not change significantly within one chunk. We use the AV1 encoder developed by AOM~\cite{chen-pcs-18} to generate the R-D points for each chunk. This is done by encoding each chunk in constant quality (CQ) mode of the AV1 encoder at $s=13$ CRF values. 
We cluster the vectors of R-D points into $10$ clusters as explained in Section~\ref{sec:clustering}.

\begin{figure}
\centering
\includegraphics[scale=0.36]{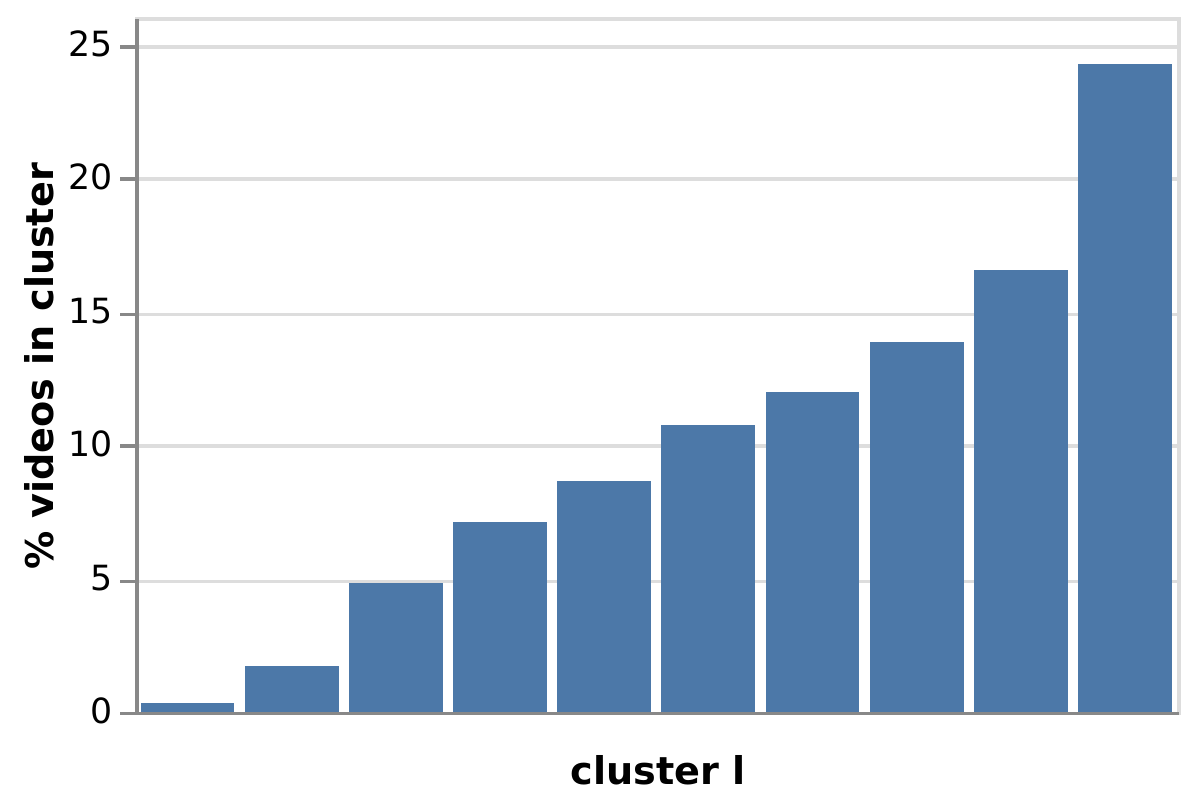}
\vspace{-0.5\baselineskip}
\caption{Distribution of number of videos in different clusters.}
\label{fig:cluster-histogram}
\vspace{-1.0\baselineskip}
\end{figure}

We train an SVM classifier for predicting the R-D cluster of a video using the features given by the first pass log of the AV1 encoder~\cite{libaom-repo}. $80\%$ of the data is used for training and the remaining $20\%$ is used for testing. We use the radial basis function (RBF) kernel in the SVM classifier. Optimal hyper-parameters of the SVM (namely, the regularization parameter and $\gamma$ used in RBF kernel~\cite{chang-svm-11}) are computed with grid search using $5$-fold cross-validation. The classifier gives an accuracy of $69\%$ on the test set.
%
%
The distribution of number of videos in different clusters (i.e., the weights $w_l$) is estimated by classifying a large sample of videos using the SVM model. It is shown in Figure~\ref{fig:cluster-histogram}.

\begin{figure}
\centering
\includegraphics[scale=0.39]{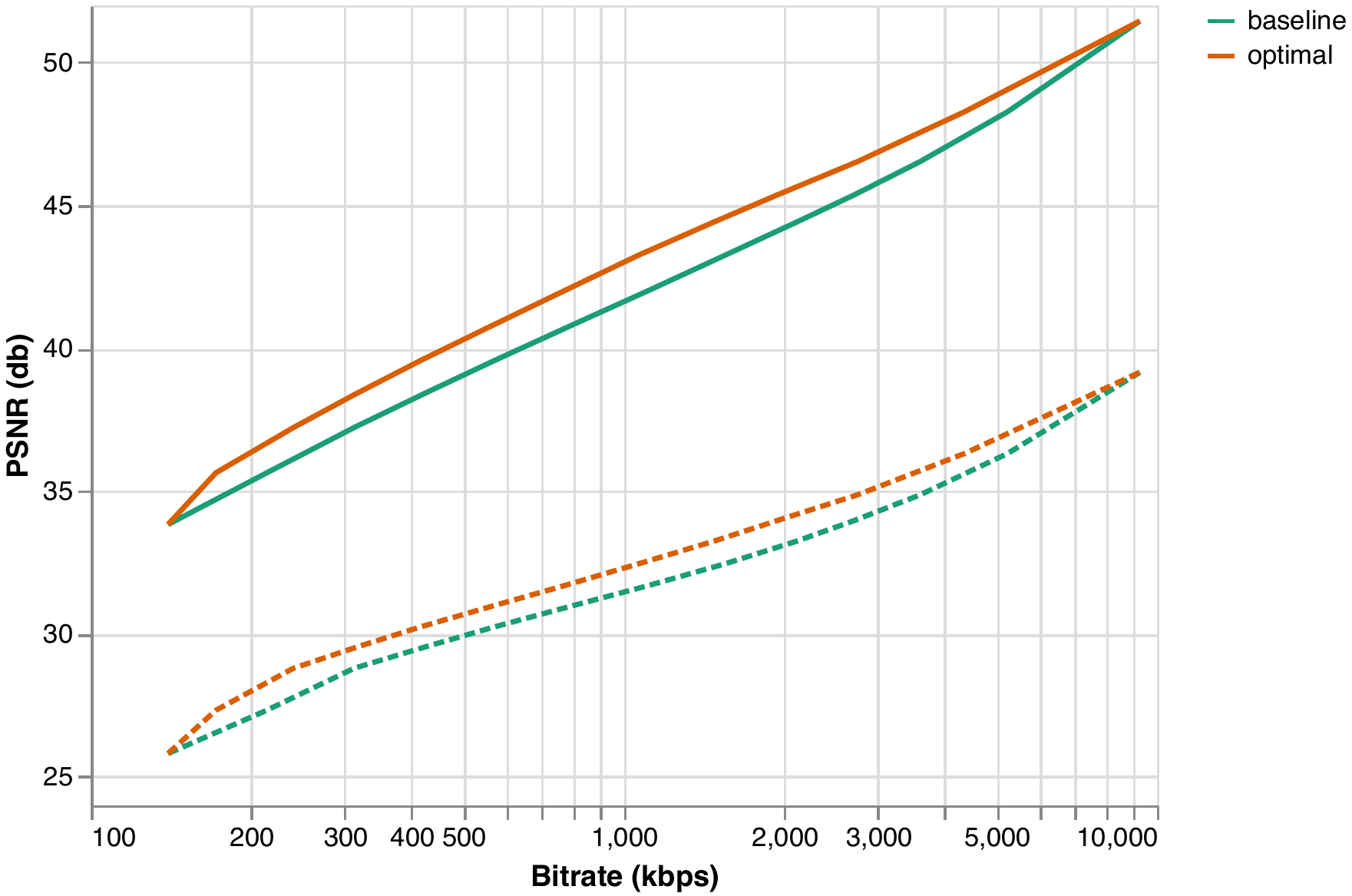}
\caption{Solid line: average PSNR vs. average bitrate. Dotted line: minimum PSNR vs. average bitrate. Computed using baseline (green) and optimal CRFs (red) with weights $w_i$ and centroid R-D curves.}
\label{fig:expected-bd-rate}
\vspace{-0.5\baselineskip}
\end{figure}

We use the weights $w_l$ and the R-D points for cluster centroids to solve optimization problem in Eq.~\eqref{eq:opt}. 
In order to get a baseline, we use the same CRF value for all R-D clusters. For each baseline CRF $q$, we compute average, $\sum_{l = 1}^{k} w_l \delta_l(q)$ and maximum, $\max_{l \in \{1, \ldots, k\}} \delta_l(q)$, distortions. We then use these values to set the constraints $D_\text{avg}$ and $D_\text{max}$ in Eq.~\eqref{eq:opt} and compute optimal CRFs, $[q_l^\star]$, for all clusters. The expected baseline and optimal average bitrates are given by $\sum_{l = 1}^{k} w_l \rho_l(q)$ and $\sum_{l = 1}^{k} w_l \rho_l(q_l^\star)$ respectively. We repeat this for multiple baseline CRF values to get baseline and optimal rate-distortion sweeps.
Figure~\ref{fig:expected-bd-rate} shows the plots of expected average and maximum distortion vs. expected average bitrate. Based on these plots, for AV1 encoder, using optimal CRF for each cluster is expected to improve the BD rate by $22\%$ compared to using the same CRF for all clusters for the same average distortion.

\begin{figure}
\centering
\includegraphics[scale=0.39]{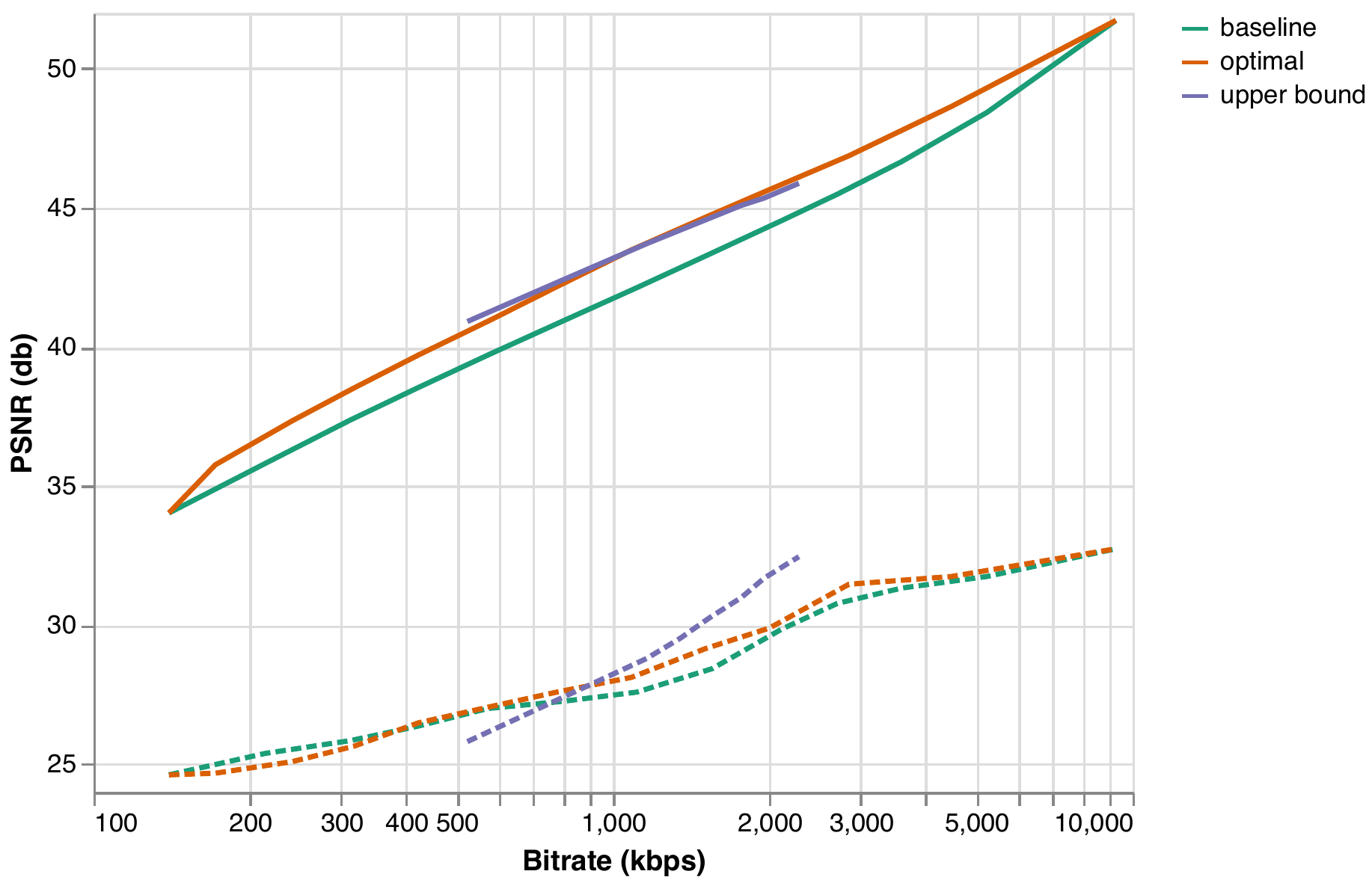}
\caption{Solid line: average PSNR vs. average bitrate. Dotted line: minimum PSNR vs. average bitrate. Computed over all chunks using optimal and baseline CRFs based on their predicted R-D class and actual R-D curves.}
\label{fig:actual-bd-rate}
\vspace{-0.6\baselineskip}
\end{figure}

We check robustness of the optimization against the inaccuracies in the R-D class prediction model and the approximation errors in centroid R-D curves. This is done by classifying each chunk in the data using the SVM classifier and then computing the bitrate and distortion at the optimal CRF for its predicted class using the actual R-D curve of the chunk. Baseline is obtained using the same CRF for all chunks. Plots of average distortion, $\frac{1}{n}\sum_{i=1}^n d_i(q_i)$, and maximum distortion, $\max_{i \in \{1, \ldots, n\}} d_i(q_i)$ vs. average bitrate, $\frac{1}{n}\sum_{i=1}^n r_i(q_i)$, computed over all chunks using optimal and baseline CRFs is shown in Figure~\ref{fig:actual-bd-rate}. These plots also show that using optimal CRFs improves the BD rate by $22\%$ compared to the baseline for the same average distortion, thus indicating that the BD rate savings persist even with modelling errors. Figure~\ref{fig:actual-bd-rate} also shows the plots of average and maximum distortions vs. average bitrate obtained by optimizing the bitrates directly for all chunks in the data (blue solid and dotted lines respectively). 
This can be considered the best rate allocation for the given chunks. 
The plots show that clustering based rate allocation achieves the smallest possible average distortion for given average bitrate. However, the maximum distortion is larger than the smallest maximum distortion possible. 
%
%

\vspace{-0.5\baselineskip}
\section{Conclusion}
\label{sec:conclusion}
\vspace{-0.5\baselineskip}
We presented an efficient method for optimal rate allocation over a large scale corpus using machine learning. Our method clusters the videos in the corpus based on their R-D characteristics and finds the optimal encoder operating points for all clusters. We developed a machine learning model to predict the R-D cluster of a test video using encoder pass log features that are easy to obtain. In the future, we would like to develop a model for finding the optimal encoder parameters using features that are even simpler to compute and reduce the amount of R-D data needed for training. It would be also interesting to integrate the playback statistics in our framework to optimize bitrates for multiple formats of the same video used for adaptive streaming.


\bibliographystyle{IEEEbib}
\bibliography{refs}

\end{document}